\documentclass[preprint2]{aastex}

\usepackage{natbib}
\usepackage{graphicx}

\newcommand{\mincir}{\raise
  -2.truept\hbox{\rlap{\hbox{$\sim$}}\raise5.truept \hbox{$<$}\ }}
\newcommand{\magcir}{\raise
  -2.truept\hbox{\rlap{\hbox{$\sim$}}\raise5.truept \hbox{$>$}\ }}

\newcommand{\eqref}[1]{(\ref{#1})}

\begin{document}

\title{Looking at the Fundamental Plane through a gravitational lens}
\author{G. Bertin and M. Lombardi\altaffilmark{1}}
\affil{%
  Universit\`a degli Studi di Milano, Dipartimento di Fisica, 
  via Celoria 16, I-20133 Milano, Italy}
\altaffiltext{1}{%
  European Southern Observatory, Karl-Schwarzschild-Stra\ss e 2,
  D-85748 Garching bei M\"unchen, Germany}

\begin{abstract}
  We consider the Fundamental Plane of elliptical galaxies lensed by
  the gravitational field of a massive deflector (typically, a cluster
  of galaxies). We show that the Fundamental Plane relation provides a
  straightforward measurement of the projected mass distribution of
  the lens with a typical accuracy of $\approx 0.15$ in the
  dimensionless column density $\kappa$.  The proposed technique
  breaks the mass-sheet degeneracy completely and is thus expected to
  serve as an important complement to other lensing-based analyses.
  Moreover, its ability to measure directly the mass distribution on
  the small pencil beams that characterize the size of background
  galaxies may lead to crucial tests for current scenarios of
  structure formation.  
\end{abstract}

\section{Introduction}
\label{sec:introduction}

One of the most remarkable properties of early-type galaxies is the
fact that their distribution in the three-dimensional parameter space
of \textrm{effective} radius $R_\mathrm{e}$, central velocity
dispersion $\sigma_0$, and average (intrinsic) surface brightness
$\langle \mathrm{SB} \rangle_\mathrm{e}$ within $R_\mathrm{e}$ is
extremely well localized.  In particular, if $\sigma_0$ is measured in
$\mbox{km s}^{-1}$, $\langle \mathrm{SB} \rangle_\mathrm{e}$ in
$\mbox{mag arcsec}^{-2}$, and $R_\mathrm{e}$ in $\mbox{kpc}$, then the
parameter space is mostly filled around a locus called the
``Fundamental Plane'' \citep{1987ApJ...313...42D, 1987ApJ...313...59D}
and described by the equation
\begin{equation}
  \label{eq:1}
  \log R_\mathrm{e} = \log r_\mathrm{e} + \log D_A(z) =
  \alpha \log \sigma_0 + \beta \bigl\langle \mathrm{SB}
  \bigr\rangle_\mathrm{e} + \gamma \; ,
\end{equation}
where $\{\alpha, \beta, \gamma\}$ are three (possibly
wavelength-dependent) constants, $r_\mathrm{e}$ is the angular
effective radius (taken to be measured in radians), and $D_A(z)$ is
the angular diameter distance of the galaxy at redshift $z$ in
$\mbox{kpc}$.  In general, the parameters $r_\mathrm{e}$ and $\langle
\mathrm{SB} \rangle_\mathrm{e} \equiv 42.0521 - 2.5 \log \bigl(L / 2
\pi R_\mathrm{e}^2 L_\odot^B \bigr)$ are operationally defined in
terms of a best fit with the $R^{1/4}$ law
\citep{1948AnAp...11..247D}; moreover, in the evaluation of the
\textit{intrinsic\/} surface brightness $\langle \mathrm{SB}
\rangle_\mathrm{e}$ from the data, a number of factors (such as
cosmological dimming, k-correction, and galactic extinction) need to
be taken into account. Note that $r_\mathrm{e}$ corresponds to the
half-light radius of the galaxy only if the luminosity profile is
exactly described by the de~Vaucouleurs law.  [Some concerns about the
universality of the $R^{1/4}$ law and the structural homology of
elliptical galaxies have been raised, but the small systematic
deviations can be interpreted in terms of weak homology, see
\citealp{2002A&A...386..149B}.]\@ Usually, the central velocity
dispersion $\sigma_0$ is defined as the luminosity weighted average
dispersion inside an aperture radius of $r_\mathrm{e} / 8$.

The tightness of the Fundamental Plane, which in the nearby universe
is characterized by a relative scatter in $R_\mathrm{e}$ of the order
of $15\%$ or below \citep[e.g.][]{1996MNRAS.280..167J}, makes this
scaling relation equivalent to the existence of a ``standard rod.''
Therefore, the Fundamental Plane has been very useful as a distance
indicator (e.g.\ \citealp{2000ApJ...529..768K,1999MNRAS.304..595G,
  1998ApJ...493..529B}; see also
\citealp{1995MNRAS.276.1255V}). Surprisingly, this scaling relation
has been found to hold tightly also at redshifts of cosmological
interest, so that it has allowed astronomers to investigate the
evolution properties of early-type galaxies out to $z \approx 1$
(\citealp{1999MNRAS.308.1037T, 1999MNRAS.308..833J,
  2003ApJ...595...29R, 2005ApJ...623..666R} and a number of related
articles; see also the discussion reported in Sect.~3 below).

So far, in the context of gravitational lensing (and mostly in the
case of strong lensing) the Fundamental Plane has been mainly employed
to characterize the properties of the deflector
\citep[see][]{2000ApJ...543..131K}, a natural approach since the
lenses are often early-type galaxies. In contrast to previous
applications, in this Letter we investigate the use of the Fundamental
Plane for the background elliptical galaxies observed through a
gravitational lens.  Since the Fundamental Plane essentially provides a
``standard rod,'' that is an \textit{absolute length-scale\/} through
$R_\mathrm{e}$, we propose to employ it to measure the magnification
due to an intervening gravitational lens.  This novel technique not
only is expected to provide measurements of the projected mass density
of the lens with relatively high signal-to-noise ratio, but also
presents several unique characteristics that make it an invaluable
diagnostics of the structure of gravitational lenses. The method is
applicable, in principle, to a study of single lensed galaxies or,
more naturally, to a small sample of lensed objects; for a given lens,
it will find best applications as a complement to other lensing-based
investigations.

\section{The Fundamental Plane of lensed galaxies}
\label{sec:fund-plane-lens}

The distortion on the observed images produced by gravitational
lensing effects is usually described in terms of the ray-tracing
function $\vec f: \vec\theta \mapsto \vec\theta^\mathrm{s}$, which
gives the \textit{real\/} position of a point source
$\vec\theta^\mathrm{s}$ given its \textit{observed\/} position
$\vec\theta$.  Gravitational lensing conserves surface brightness
\citep{SEF}, and consequently the observed intensity $I$ is related to
the true, unlensed intensity $I^\mathrm{s}$ by the simple equation
\begin{equation}
  \label{eq:2}
  I(\vec\theta) = I^\mathrm{s}(\vec\theta^\mathrm{s}) =
  I^\mathrm{s}\bigl(\vec f(\vec\theta)\bigr) \; .
\end{equation}
If the (projected) lens mass distribution is smooth on small scales
(for example, on the typical angular size of the images of background
galaxies), the ray-tracing function will also be smooth on these
scales.  As a result, it is possible to Taylor expand to first order
this function as
\begin{eqnarray}
  \label{eq:3}
  \vec f(\vec\theta_0 + \vec\delta) = {} & \vec f(\vec\theta_0) + \left
    . \frac{\partial\vec f}{\partial\vec \theta}
    \right|_{\vec\theta_0} \vec\delta + O(\delta^2) \; \\
    {} \equiv {} & \vec\theta^\mathrm{s}_0 + A \vec\delta +
    O(\delta^2) \; ,
\end{eqnarray}
where the last line defines the source position
$\vec\theta^\mathrm{s}_0 \equiv f(\vec\theta_0)$ and the Jacobian of
the ray-tracing $A$.  In this approximation, the quantities
entering the Fundamental Plane are modified by the ray-tracing
function in the following way:
\begin{eqnarray}
  \label{eq:4}
  r_\mathrm{e} & {} \mapsto r_\mathrm{e}^\mathrm{s} =  r_\mathrm{e}
  \sqrt{\bigl| \det A \bigr|} \; , \\
  \label{eq:5}
  \sigma_0 & {} \mapsto \sigma_0^\mathrm{s} =
  \sigma_0 \; , \\
  \label{eq:6}
  \bigl\langle \mathrm{SB} \bigr\rangle_\mathrm{e} & {} \mapsto
  \bigl\langle \mathrm{SB} \bigr\rangle_\mathrm{e}^\mathrm{s} =
  \bigl\langle \mathrm{SB} \bigr\rangle_\mathrm{e} \; .
\end{eqnarray}
The last two equations hold because both the point values of the
velocity dispersion and of the surface brightness are conserved by
gravitational lensing, and because both values are defined
\textit{intrinsically\/} with respect to the effective radius
$r_\mathrm{e}$.  We note that the simple transformation \eqref{eq:4}
is valid only when the effective \textit{circularized\/} radius
$r_\mathrm{e}$ is defined (as is usually done) as the geometric mean
of the galaxy semiaxes $r_\mathrm{e} = \sqrt{a b}$.  In conclusion,
Eq.~\eqref{eq:1} refers to the \textit{unlensed\/} quantities $\bigr\{
r^\mathrm{s}_\mathrm{e}, \sigma^\mathrm{s}_0, \langle \mathrm{SB}
\rangle_\mathrm{e} \bigr\}$. As a result, in the presence of
gravitational lensing, the Fundamental Plane is transformed into
\begin{equation}
  \label{eq:7}
  \log r_\mathrm{e} + \log D_A(z) =
  \alpha \log \sigma_0 + \beta \bigl\langle \mathrm{SB}
  \bigr\rangle_\mathrm{e} + \gamma - \frac{1}{2} \log \, | \det A
  | \; .
\end{equation}
In this equation, all terms but the last can be derived from the
observations. The Fundamental Plane relation can thus be inverted
to obtain $| \det A |$.

\section{Discussion}
\label{sec:discussion}

Similarly to the case of the so-called magnification effect
\citep[e.g.][]{1998ApJ...501..539T}, our technique produces an
estimate of the determinant of the Jacobian matrix $A$ of the
ray-tracing. In the case of a single-screen lens, we have
\citep[see][]{SEF} $\det A = (1 - \kappa)^2 (1 - g)^2$, where $\kappa
= \Sigma / \Sigma_\mathrm{c}$ is the lens \textit{convergence\/}
(i.e.\ the ratio between the lens projected mass distribution $\Sigma$
and the geometrical critical density $\Sigma_\mathrm{c}$), and $g$ is
the modulus of the \textit{reduced\/} shear (the ratio between the
shear and the quantity $1 - \kappa$). In the weak lensing limit (for
which the present technique is more straightforward) both $\kappa$ and
$g$ are small quantities and the determinant reduces simply to
$\det A \approx 1 - 2 \kappa$. Hence, in this important limit our
technique provides directly a measurement of the lens projected mass
density:
\begin{equation}
  \label{eq:8}
  0.4343 \, \kappa \approx \log r_\mathrm{e} + \log D_A(z) -
  \alpha \log \sigma_0 - \beta \langle \mathrm{SB} \rangle_\mathrm{e}
  - \gamma \; .
\end{equation}

Before considering the advantages and the limitations of the method
proposed in this Letter, we should address the issue of the
statistical error associated with a single application of
Eq.~\eqref{eq:8}. Several studies of the nearby universe have shown
that the Fundamental Plane is extremely tight. In some cases it has
been claimed that the intrinsic scatter in $R_\mathrm{e}$ is as low as
$11\%$ \citep{1993ApJ...411...34J}; more recent investigations find
that the intrinsic scatter of the Fundamental Plane is approximately
$15\%$ in $R_\mathrm{e}$ \citep{1996MNRAS.280..167J}. Surprisingly,
the scatter appears to be very small also for samples of galaxies at
relatively high redshift \citep{2003AJ....125.1866B,
  2005A&A...442..125D}. If we refer to the more conservative relative
error of $15\%$ in $R_\mathrm{e}$, we derive a relative error of
$30\%$ in $| \det A |$ and thus an (absolute) error of
$\approx 0.15$ in $\kappa$.  \textit{Hence, the effect considered in
  this paper should be easily detected from a single observation in
  the central regions of massive clusters\/}; alternatively, one could
use collectively several observations in the periphery of clusters
(see further description below at the end of this section).

Another issue to be addressed is related to the evolution of the
Fundamental Plane. So far no clear indication of a redshift dependence
of the constants $\alpha$ and $\beta$ in Eq.~\eqref{eq:1} has been
found, while the offset of the plane $\gamma$ depends on redshift and
is typically consistent with a picture of passive galaxy
evolution. These features can be easily incorporated in our study if
we have good empirical knowledge of $\gamma(z)$ for field galaxies
(e.g., see \citealp{2002ApJ...564L..13T}; but see the discussion
below).

The lensing of the Fundamental Plane presents several advantages
when compared to standard gravitational lensing techniques:
\begin{itemize}
\item It measures directly the projected mass density and not the
  shear, as typically done in weak lensing.  As a result, the
  measurements can be interpreted in terms of clear physical quantities
  with no need of any further analysis.
\item Our method is not plagued by the mass-sheet degeneracy, which
  severely hampers lensing studies \citep[see, e.g.][]{RevBS,
    2004A&A...424...13B}. A similar advantage would be shared by the
  magnification effect studied previously \citep{1998ApJ...501..539T},
  but its application has met major difficulties, mainly because of
  its sensitivity to spurious changes in the observed density of
  background galaxies or to inaccurate measurements of the unlensed
  galaxy density \citep{2000A&A...353...41S}.  In addition, a severe
  bias is generally introduced by the presence of bright galaxies,
  which tend to saturate the central regions of clusters and to
  significantly affect the completeness in the detection of background
  galaxies. In contrast, the lensing of the Fundamental Plane is not
  affected by any apparent (position-dependent) bias and should thus
  produce reliable measurements.
\item The technique proposed yields measurements characterized by a
  high signal-to-noise ratio, with a typical error on $\kappa$ of
  $\approx 0.15$.  By comparison, we note that a single galaxy usually
  provides a shear measurement with an accuracy of $\approx 0.3$ or
  worse \citep[see, e.g.][]{2002AJ....123..583B}. For a typical lens
  $g \sim \kappa$, i.e.\ the reduced shear modulus and the convergence
  have the same order of magnitude, but since the shear depends
  non-locally on the lens mass distribution, several thousands of
  independent shear measurements are usually needed to obtain a good
  weak lensing detection \citep[e.g.][]{2005ApJ...623...42L}. In this
  respect, a small number of ``lensed Fundamental Plane'' measurements
  can be used together with standard lensing analyses as an efficient
  way to break the mass-sheet degeneracy.
\item The lensing of the Fundamental Plane is based on the evaluation
  of the lensing magnification on small pencil beams corresponding to
  the typical size of a background galaxy (usually a few arcseconds).
  These high-resolution measurements are severely undersampled, and
  will need to be interpolated to obtain a smooth map of the lens mass
  distribution. Yet, the ability to obtain direct, reliable estimates
  of the projected mass distribution on small angular scales is bound
  to offer an invaluable diagnostics to test current structure
  formation scenarios (for example, to confirm the existence of small
  undetected dark satellites around massive galaxies, see
  \citealp{1993MNRAS.264..201K, 1999ApJ...524L..19M,
    1999ApJ...522...82K} or of dark density clumps in clusters of
  galaxies).
\end{itemize}
The merits described above come with some price to pay.  First, we
note that the study of the Fundamental Plane at intermediate and high
redshifts is challenging, especially for the low-density environments
considered in this Letter.  In particular, the precise evolution of
$\gamma$ with redshift, $\gamma(z)$, is still under debate.  For
example, \citet{2000ApJ...543..131K}, \citet{2001ApJ...553L..39V},
\citet{2003ApJ...592L..53V}, and \citet{2003ApJ...587..143R} report a
relatively slow evolution of $\gamma(z)$, while
\citet{2002ApJ...564L..13T} and \citet{2003ApJ...597..239G} claim a
somewhat faster evolution.  Unfortunately, an error in the estimate of
the Fundamental Plane intercept $\gamma(z)$ will directly affect the
estimate of the lensing magnification, and thus of the lens
dimensionless mass density $\kappa$.  Currently, the determination of
the derivative of $\gamma$ with redshift ranges from $\gamma'(z) =
1.35^{+0.2}_{-0.3}$ \citep{2002ApJ...564L..13T} to $\gamma'(z) = 1.00
\pm 0.12$ \citep{2003ApJ...592L..53V}: hence, these two extremes would
lead to a systematic uncertainty of $\approx 0.1$ for $\gamma$ at $z
\approx 0.8$.  On the other hand, since several independent studies
(including the ones mentioned above) have shown that the thickness of
the Fundamental Plane does not increase significantly with redshift,
the differences reported must be related to systematic effects rather
than to pure statistical uncertainties.  As a result, it is natural to
expect that in the near future it will be possible to reduce this
error source and thus to base the study that we propose on a well
reliable estimate of $\gamma(z)$ for field galaxies.

% beta = 0.32  ==>  Delta gamma = - Delta ln M/L_b * 0.8
%\citep{2001ApJ...553L..39V}: Delta ln M/L_B = (-1.35 +/- 0.35) z
%\citep{2003ApJ...592L..53V}: Delta ln M/L_B = (-1.25 +/- 0.15) z
%\citep{2001MNRAS.326..237T}: ?
%\citep{2002ApJ...564L..13T}: Delta ln M/L_B = (-1.68 + 0.25 - 0.37) z

The point just discussed remind us that the acquisition of the
relevant Fundamental Plane data for distant early-type galaxies is
non-trivial.  Therefore, observations will necessarily be limited to
relatively bright objects (currently $R < 23 \mbox{ mag}$).  In
addition, elliptical galaxies represent only a fraction (approximately
$30\%$) of all galaxies, and their relative number decreases with
redshift. In conclusion, all these difficulties suggest that the
present application of the method proposed here should be viable only
on low-to-mid redshift lensing clusters of galaxies, for which the
probability of identifying background early-type galaxies in the field
is sufficiently high.

In order to address these difficulties more quantitatively, we first
note that, as pointed out by various authors (e.g.,
\citealp{2000ApJ...531..137K}), the combination of parameters that
enters the Fundamental Plane is particularly robust because it is
characterized by (anti-)correlated errors.  In particular, for the
sample of mid-redshift galaxies studied by
\citet{2001MNRAS.326..237T}, the median measurement error on the
combination $\log R_\mathrm{e} - \beta \langle SB \rangle_\mathrm{e}$
is as small as $0.02$.  A larger contribution to the scatter of the
Fundamental Plane is given by the velocity dispersion measurements.
For example, if we conservatively assume that measurements on $\sigma$
are performed with a $\approx 10\%$ accuracy, then we expect a scatter
of approximately $0.05$ in $\log R_\mathrm{e}$, to be compared with
the $\approx 0.07$ intrinsic scatter in $\log R_\mathrm{e}$.  In summary,
if we take into account typical measurement errors on the Fundamental
Plane, the expected error on $R_\mathsf{e}$ may increase from $\approx
15\%$ to $\approx 20\%$.

Another important point to be addressed in detail is the expected
angular density of ellipticals for which we can estimate the
parameters entering the Fundamental Plane with reasonable accuracy.
Current 8-meter class telescopes can obtain high signal-to-noise
spectra of objects as faint as $21.5 \mbox{ mag}$ in the $z$-SDSS band
\citep{2005ApJ...631..145V}, corresponding to $\mathrm{F814W} \approx
21.7 \mbox{ mag}$ and $R \approx 23.3 \mbox{ mag}$ (the color
conversions are according to \citealp{1995PASP..107..945F}).  Given
the number counts of elliptical galaxies \citep[e.g.,
see][]{1995MNRAS.275L..19G}, we estimate that there are approximately
$2 \mbox{ galaxies arcmin}^{-2}$ brighter than $21.5 \mbox{ mag}$ in
$z$-SDSS.  Assuming, for simplicity, that the cluster mass distribution
is approximately described by a $\approx 1000 \mbox{ km s}^{-1}$ singular
isothermal sphere, we estimate that within a $1 \mbox{ arcmin}$
radius, the reduced mass density $\kappa$ exceeds $0.15$ for a source
at redshift $0.8$.  In other words, for such a lens we would expect
approximately $5$ background early-type galaxies showing (each of them
\textit{individually\/}) detectable lensing of the Fundamental Plane.
In reality, this number is likely to be underestimated, also because
clusters have often a complex morphology and the presence of
substructure usually increases the useful area where lensing
investigations can be carried out.  For example, in Abell~1689 at $z =
0.18$, the estimated dimensionless mass density for a source at
redshift $0.8$ exceed $0.5$ within a $1 \mbox{ arcmin}$ radius, and
decreases below $0.15$ only beyond the $2 \mbox{ arcmin}$ radius
\citep{2005ApJ...621...53B, 2005MNRAS.362.1247D}.

\section{Conclusions and prospects}
\label{sec:conclusions}

In this Letter we have considered the application of the Fundamental
Plane of elliptical galaxies, as equivalent to a ``standard rod'', to
investigate the mass distribution of gravitational lenses from their
magnification of the intrinsic scale-length $R_\mathrm{e}$. We have
shown that the proposed technique can provide column density
measurements on pencil beams with an accuracy of $0.15$ in the reduced
dimensionless mass density $\kappa$.

As explained above, our technique has several advantages with respect
to standard lensing methods, but is also observationally challenging
and can be applied to the limited number of early-type galaxies
typically observed through a lens.  These considerations suggest that
the lensing of the Fundamental Plane would be best employed in
conjunction with other lensing analyses.  In fact, ground-based weak
lensing observations are based on approximately $25 \mbox{ galaxies
  arcmin}^{-2}$, with each galaxy typically giving an estimate of the
local shear with $0.4$ error.  Hence, by averaging the various shear
measurements over a square arcmin, an error of $\approx 0.08$ on the
shear can be achieved.  This value should be compared with the $\approx
0.1$ error expected on the convergence $\kappa$ from our method alone,
or with the $\approx 0.06$ error from a combined analysis.  In addition,
as noted earlier in this Letter, a combined analysis would break the
mass-sheet degeneracy of weak lensing.

In principle, an analogous technique could be based on the
existence of a ``standard candle'', as provided by the Tully-Fisher
\citep{1977A&A....54..661T} relation for late-type galaxies,
\begin{equation}
  \label{eq:9}
  L \propto V^p \; ,
\end{equation}
where $L$ is the total absolute luminosity of the galaxy, $V$ is its
\textit{maximum\/} rotational velocity, and $p$ is a
(wavelength-dependent) constant.  Similarly to the Fundamental Plane,
the Tully-Fisher scaling relation is often used as a distance
estimator or to analyse the evolution properties of spiral galaxies.
In the nearby universe, the measured scatter in $L$ can be as small as
$20\%$ (e.g.\ \citealp{1988ApJ...330..579P, 1990ApJ...351L...5W,
  1997AJ....113.2046R, 2001ApJ...563..694V}; see
\citealp{1995PhR...261..271S} for a review), suggesting that the
Tully-Fisher relation might be used to measure with a similar accuracy
the lens magnification $| \det A |$, and would thus provide a direct
estimate of $\kappa$ with an rms error as small as $0.1$.  The
practical use of the Tully-Fisher relation as a distance estimator has
been limited to low redshifts.  Recently, it has been possible to
investigate the Tully-Fisher relation at relatively high redshifts
\citep[e.g.][]{2006MNRAS.366..308B, 2004A&A...420...97B}.
Unfortunately, a rather large scatter is observed in these studies
(approximately $0.8$--$1 \mbox{ mag}$), which severely limits
applications of the type proposed in this Letter on the basis of the
Fundamental Plane. Nevertheless, if the observed scatter is not
(totally) intrinsic, next generation telescopes, such as the James
Webb Space Telescope or the Extremely Large Telescopes, may open the
way to this other, possibly powerful, investigation technique.

Type Ia supernovae (SNe Ia) also appear to be very good (calibrated)
``standard candles'' \citep{1996ApJ...473...88R, 1998ARA&A..36...17B}
and thus they could be suitable for the lensing analysis proposed in
this Letter.  In fact, such a possibility has already been considered
by various authors (see, e.g., \citealp{2001ApJ...556L..71H,
  2002A&A...393...25G, 2003ApJ...583..584O, 2003MNRAS.338L..25O}),
especially in relation to the prospects offered by the
\textit{Supernova/Acceleration Probe\/} (\textit{SNAP\/}).  This
satellite is expected to provide $\approx 8$ multiple imaged
supernovae per year within its 20 fixed, $1 \mbox{ deg}^2$ fields
\citep{2001ApJ...556L..71H}; of these, $\approx 2$ are expected to be
Type Ia SNe for which the \textit{absolute\/} magnification can be
measured.  Since the probability of observing a strong lensing effect
on a randomly selected area of the sky (the strong lensing optical
depth) is $\tau \approx 0.001$ for sources out to redshift $1.5$,
lensing applications of SNe will be limited to the study of the few
clusters for which serendipitous observations become available.

\acknowledgements We wish to thank Piero Rosati for several
helpful and stimulating discussions.  This work was partly supported
by MIUR (Cofin-2004).

\bibliographystyle{aa}
\bibliography{../lens-refs.bib,../dyn-refs.bib}

\begin{thebibliography}{53}
\expandafter\ifx\csname natexlab\endcsname\relax\def\natexlab#1{#1}\fi

\bibitem[{{Bamford} {et~al.}(2006){Bamford}, {Arag{\'o}n-Salamanca}, \&
  {Milvang-Jensen}}]{2006MNRAS.366..308B}
{Bamford}, S.~P., {Arag{\'o}n-Salamanca}, A., \& {Milvang-Jensen}, B. 2006,
  \mnras, 366, 308

\bibitem[{{Bartelmann} \& {Schneider}(2001)}]{RevBS}
{Bartelmann}, M. \& {Schneider}, P. 2001, {Physics} {Reports}, 340, 291

\bibitem[{{Bender} {et~al.}(1998){Bender}, {Saglia}, {Ziegler}, {Belloni},
  {Greggio}, {Hopp}, \& {Bruzual}}]{1998ApJ...493..529B}
{Bender}, R., {Saglia}, R.~P., {Ziegler}, B., {et~al.} 1998, \apj, 493, 529

\bibitem[{{Bernardi} {et~al.}(2003){Bernardi}, {Sheth}, {Annis}, {Burles},
  {Eisenstein}, {Finkbeiner}, {Hogg}, {Lupton}, {Schlegel}, {SubbaRao},
  {Bahcall}, {Blakeslee}, {Brinkmann}, {Castander}, {Connolly}, {Csabai},
  {Doi}, {Fukugita}, {Frieman}, {Heckman}, {Hennessy}, {Ivezi{\'c}}, {Knapp},
  {Lamb}, {McKay}, {Munn}, {Nichol}, {Okamura}, {Schneider}, {Thakar}, \&
  {York}}]{2003AJ....125.1866B}
{Bernardi}, M., {Sheth}, R.~K., {Annis}, J., {et~al.} 2003, \aj, 125, 1866

\bibitem[{{Bernstein} \& {Jarvis}(2002)}]{2002AJ....123..583B}
{Bernstein}, G.~M. \& {Jarvis}, M. 2002, \aj, 123, 583

\bibitem[{{Bertin} {et~al.}(2002){Bertin}, {Ciotti}, \& {Del
  Principe}}]{2002A&A...386..149B}
{Bertin}, G., {Ciotti}, L., \& {Del Principe}, M. 2002, \aap, 386, 149

\bibitem[{{B{\"o}hm} {et~al.}(2004){B{\"o}hm}, {Ziegler}, {Saglia}, {Bender},
  {Fricke}, {Gabasch}, {Heidt}, {Mehlert}, {Noll}, \&
  {Seitz}}]{2004A&A...420...97B}
{B{\"o}hm}, A., {Ziegler}, B.~L., {Saglia}, R.~P., {et~al.} 2004, \aap, 420, 97

\bibitem[{{Brada{\v c}} {et~al.}(2004){Brada{\v c}}, {Lombardi}, \&
  {Schneider}}]{2004A&A...424...13B}
{Brada{\v c}}, M., {Lombardi}, M., \& {Schneider}, P. 2004, \aap, 424, 13

\bibitem[{{Branch}(1998)}]{1998ARA&A..36...17B}
{Branch}, D. 1998, \araa, 36, 17

\bibitem[{{Broadhurst} {et~al.}(2005){Broadhurst}, {Ben{\'{\i}}tez}, {Coe},
  {Sharon}, {Zekser}, {White}, {Ford}, {Bouwens}, {Blakeslee}, {Clampin},
  {Cross}, {Franx}, {Frye}, {Hartig}, {Illingworth}, {Infante}, {Menanteau},
  {Meurer}, {Postman}, {Ardila}, {Bartko}, {Brown}, {Burrows}, {Cheng},
  {Feldman}, {Golimowski}, {Goto}, {Gronwall}, {Herranz}, {Holden}, {Homeier},
  {Krist}, {Lesser}, {Martel}, {Miley}, {Rosati}, {Sirianni}, {Sparks},
  {Steindling}, {Tran}, {Tsvetanov}, \& {Zheng}}]{2005ApJ...621...53B}
{Broadhurst}, T., {Ben{\'{\i}}tez}, N., {Coe}, D., {et~al.} 2005, \apj, 621, 53

\bibitem[{{de Vaucouleurs}(1948)}]{1948AnAp...11..247D}
{de Vaucouleurs}, G. 1948, Annales d'Astrophysique, 11, 247

\bibitem[{{di Serego Alighieri} {et~al.}(2005){di Serego Alighieri}, {Vernet},
  {Cimatti}, {Lanzoni}, {Cassata}, {Ciotti}, {Daddi}, {Mignoli}, {Pignatelli},
  {Pozzetti}, {Renzini}, {Rettura}, \& {Zamorani}}]{2005A&A...442..125D}
{di Serego Alighieri}, S., {Vernet}, J., {Cimatti}, A., {et~al.} 2005, \aap,
  442, 125

\bibitem[{{Diego} {et~al.}(2005){Diego}, {Sandvik}, {Protopapas}, {Tegmark},
  {Ben{\'{\i}}tez}, \& {Broadhurst}}]{2005MNRAS.362.1247D}
{Diego}, J.~M., {Sandvik}, H.~B., {Protopapas}, P., {et~al.} 2005, \mnras, 362,
  1247

\bibitem[{{Djorgovski} \& {Davis}(1987)}]{1987ApJ...313...59D}
{Djorgovski}, S. \& {Davis}, M. 1987, \apj, 313, 59

\bibitem[{{Dressler} {et~al.}(1987){Dressler}, {Lynden-Bell}, {Burstein},
  {Davies}, {Faber}, {Terlevich}, \& {Wegner}}]{1987ApJ...313...42D}
{Dressler}, A., {Lynden-Bell}, D., {Burstein}, D., {et~al.} 1987, \apj, 313, 42

\bibitem[{{Fukugita} {et~al.}(1995){Fukugita}, {Shimasaku}, \&
  {Ichikawa}}]{1995PASP..107..945F}
{Fukugita}, M., {Shimasaku}, K., \& {Ichikawa}, T. 1995, \pasp, 107, 945

\bibitem[{{Gavazzi} {et~al.}(1999){Gavazzi}, {Boselli}, {Scodeggio}, {Pierini},
  \& {Belsole}}]{1999MNRAS.304..595G}
{Gavazzi}, G., {Boselli}, A., {Scodeggio}, M., {Pierini}, D., \& {Belsole}, E.
  1999, \mnras, 304, 595

\bibitem[{{Gebhardt} {et~al.}(2003){Gebhardt}, {Faber}, {Koo}, {Im}, {Simard},
  {Illingworth}, {Phillips}, {Sarajedini}, {Vogt}, {Weiner}, \&
  {Willmer}}]{2003ApJ...597..239G}
{Gebhardt}, K., {Faber}, S.~M., {Koo}, D.~C., {et~al.} 2003, \apj, 597, 239

\bibitem[{{Glazebrook} {et~al.}(1995){Glazebrook}, {Ellis}, {Santiago}, \&
  {Griffiths}}]{1995MNRAS.275L..19G}
{Glazebrook}, K., {Ellis}, R., {Santiago}, B., \& {Griffiths}, R. 1995, \mnras,
  275, L19

\bibitem[{{Goobar} {et~al.}(2002){Goobar}, {M{\"o}rtsell}, {Amanullah}, \&
  {Nugent}}]{2002A&A...393...25G}
{Goobar}, A., {M{\"o}rtsell}, E., {Amanullah}, R., \& {Nugent}, P. 2002, \aap,
  393, 25

\bibitem[{{Holz}(2001)}]{2001ApJ...556L..71H}
{Holz}, D.~E. 2001, \apjl, 556, L71

\bibitem[{{J{\o}rgensen} {et~al.}(1999){J{\o}rgensen}, {Franx}, {Hjorth}, \&
  {van Dokkum}}]{1999MNRAS.308..833J}
{J{\o}rgensen}, I., {Franx}, M., {Hjorth}, J., \& {van Dokkum}, P.~G. 1999,
  \mnras, 308, 833

\bibitem[{{J{\o}rgensen} {et~al.}(1993){J{\o}rgensen}, {Franx}, \&
  {Kjaergaard}}]{1993ApJ...411...34J}
{J{\o}rgensen}, I., {Franx}, M., \& {Kjaergaard}, P. 1993, \apj, 411, 34

\bibitem[{{J{\o}rgensen} {et~al.}(1996){J{\o}rgensen}, {Franx}, \&
  {Kjaergaard}}]{1996MNRAS.280..167J}
{J{\o}rgensen}, I., {Franx}, M., \& {Kjaergaard}, P. 1996, \mnras, 280, 167

\bibitem[{{Kauffmann} {et~al.}(1993){Kauffmann}, {White}, \&
  {Guiderdoni}}]{1993MNRAS.264..201K}
{Kauffmann}, G., {White}, S.~D.~M., \& {Guiderdoni}, B. 1993, \mnras, 264, 201

\bibitem[{{Kelson} {et~al.}(2000{\natexlab{a}}){Kelson}, {Illingworth},
  {Tonry}, {Freedman}, {Kennicutt}, {Mould}, {Graham}, {Huchra}, {Macri},
  {Madore}, {Ferrarese}, {Gibson}, {Sakai}, {Stetson}, {Ajhar}, {Blakeslee},
  {Dressler}, {Ford}, {Hughes}, {Sebo}, \& {Silbermann}}]{2000ApJ...529..768K}
{Kelson}, D.~D., {Illingworth}, G.~D., {Tonry}, J.~L., {et~al.}
  2000{\natexlab{a}}, \apj, 529, 768

\bibitem[{{Kelson} {et~al.}(2000{\natexlab{b}}){Kelson}, {Illingworth}, {van
  Dokkum}, \& {Franx}}]{2000ApJ...531..137K}
{Kelson}, D.~D., {Illingworth}, G.~D., {van Dokkum}, P.~G., \& {Franx}, M.
  2000{\natexlab{b}}, \apj, 531, 137

\bibitem[{{Klypin} {et~al.}(1999){Klypin}, {Kravtsov}, {Valenzuela}, \&
  {Prada}}]{1999ApJ...522...82K}
{Klypin}, A., {Kravtsov}, A.~V., {Valenzuela}, O., \& {Prada}, F. 1999, \apj,
  522, 82

\bibitem[{{Kochanek} {et~al.}(2000){Kochanek}, {Falco}, {Impey}, {Leh{\'a}r},
  {McLeod}, {Rix}, {Keeton}, {Mu{\~n}oz}, \& {Peng}}]{2000ApJ...543..131K}
{Kochanek}, C.~S., {Falco}, E.~E., {Impey}, C.~D., {et~al.} 2000, \apj, 543,
  131

\bibitem[{{Lombardi} {et~al.}(2005){Lombardi}, {Rosati}, {Blakeslee}, {Ettori},
  {Demarco}, {Ford}, {Illingworth}, {Clampin}, {Hartig}, {Ben{\'{\i}}tez},
  {Broadhurst}, {Franx}, {Jee}, {Postman}, \& {White}}]{2005ApJ...623...42L}
{Lombardi}, M., {Rosati}, P., {Blakeslee}, J.~P., {et~al.} 2005, \apj, 623, 42

\bibitem[{{Moore} {et~al.}(1999){Moore}, {Ghigna}, {Governato}, {Lake},
  {Quinn}, {Stadel}, \& {Tozzi}}]{1999ApJ...524L..19M}
{Moore}, B., {Ghigna}, S., {Governato}, F., {et~al.} 1999, \apjl, 524, L19

\bibitem[{{Oguri} \& {Kawano}(2003)}]{2003MNRAS.338L..25O}
{Oguri}, M. \& {Kawano}, Y. 2003, \mnras, 338, L25

\bibitem[{{Oguri} {et~al.}(2003){Oguri}, {Suto}, \&
  {Turner}}]{2003ApJ...583..584O}
{Oguri}, M., {Suto}, Y., \& {Turner}, E.~L. 2003, \apj, 583, 584

\bibitem[{{Pierce} \& {Tully}(1988)}]{1988ApJ...330..579P}
{Pierce}, M.~J. \& {Tully}, R.~B. 1988, \apj, 330, 579

\bibitem[{{Raychaudhury} {et~al.}(1997){Raychaudhury}, {von Braun},
  {Bernstein}, \& {Guhathakurta}}]{1997AJ....113.2046R}
{Raychaudhury}, S., {von Braun}, K., {Bernstein}, G.~M., \& {Guhathakurta}, P.
  1997, \aj, 113, 2046

\bibitem[{{Riess} {et~al.}(1996){Riess}, {Press}, \&
  {Kirshner}}]{1996ApJ...473...88R}
{Riess}, A.~G., {Press}, W.~H., \& {Kirshner}, R.~P. 1996, \apj, 473, 88

\bibitem[{{Rusin} \& {Kochanek}(2005)}]{2005ApJ...623..666R}
{Rusin}, D. \& {Kochanek}, C.~S. 2005, \apj, 623, 666

\bibitem[{{Rusin} {et~al.}(2003{\natexlab{a}}){Rusin}, {Kochanek}, {Falco},
  {Keeton}, {McLeod}, {Impey}, {Leh{\'a}r}, {Mu{\~n}oz}, {Peng}, \&
  {Rix}}]{2003ApJ...587..143R}
{Rusin}, D., {Kochanek}, C.~S., {Falco}, E.~E., {et~al.} 2003{\natexlab{a}},
  \apj, 587, 143

\bibitem[{{Rusin} {et~al.}(2003{\natexlab{b}}){Rusin}, {Kochanek}, \&
  {Keeton}}]{2003ApJ...595...29R}
{Rusin}, D., {Kochanek}, C.~S., \& {Keeton}, C.~R. 2003{\natexlab{b}}, \apj,
  595, 29

\bibitem[{Schneider {et~al.}(1992)Schneider, Ehlers, \& Falco}]{SEF}
Schneider, P., Ehlers, J., \& Falco, E.~E. 1992, Gravitational Lenses (Berlin
  Heidelberk NewYork: {Springer-Verlag})

\bibitem[{{Schneider} {et~al.}(2000){Schneider}, {King}, \&
  {Erben}}]{2000A&A...353...41S}
{Schneider}, P., {King}, L., \& {Erben}, T. 2000, \aap, 353, 41

\bibitem[{{Strauss} \& {Willick}(1995)}]{1995PhR...261..271S}
{Strauss}, M.~A. \& {Willick}, J.~A. 1995, \physrep, 261, 271

\bibitem[{{Taylor} {et~al.}(1998){Taylor}, {Dye}, {Broadhurst}, {Benitez}, \&
  {Van Kampen}}]{1998ApJ...501..539T}
{Taylor}, A.~N., {Dye}, S., {Broadhurst}, T.~J., {Benitez}, N., \& {Van
  Kampen}, E. 1998, \apj, 501, 539

\bibitem[{{Treu} {et~al.}(2001){Treu}, {Stiavelli}, {Bertin}, {Casertano}, \&
  {M{\o}ller}}]{2001MNRAS.326..237T}
{Treu}, T., {Stiavelli}, M., {Bertin}, G., {Casertano}, S., \& {M{\o}ller}, P.
  2001, \mnras, 326, 237

\bibitem[{{Treu} {et~al.}(1999){Treu}, {Stiavelli}, {Casertano}, {M{\o}ller},
  \& {Bertin}}]{1999MNRAS.308.1037T}
{Treu}, T., {Stiavelli}, M., {Casertano}, S., {M{\o}ller}, P., \& {Bertin}, G.
  1999, \mnras, 308, 1037

\bibitem[{{Treu} {et~al.}(2002){Treu}, {Stiavelli}, {Casertano}, {M{\o}ller},
  \& {Bertin}}]{2002ApJ...564L..13T}
{Treu}, T., {Stiavelli}, M., {Casertano}, S., {M{\o}ller}, P., \& {Bertin}, G.
  2002, \apjl, 564, L13

\bibitem[{{Tully} \& {Fisher}(1977)}]{1977A&A....54..661T}
{Tully}, R.~B. \& {Fisher}, J.~R. 1977, \aap, 54, 661

\bibitem[{{van Albada} {et~al.}(1995){van Albada}, {Bertin}, \&
  {Stiavelli}}]{1995MNRAS.276.1255V}
{van Albada}, T.~S., {Bertin}, G., \& {Stiavelli}, M. 1995, \mnras, 276, 1255

\bibitem[{{van der Wel} {et~al.}(2005){van der Wel}, {Franx}, {van Dokkum},
  {Rix}, {Illingworth}, \& {Rosati}}]{2005ApJ...631..145V}
{van der Wel}, A., {Franx}, M., {van Dokkum}, P.~G., {et~al.} 2005, \apj, 631,
  145

\bibitem[{{van Dokkum} \& {Ellis}(2003)}]{2003ApJ...592L..53V}
{van Dokkum}, P.~G. \& {Ellis}, R.~S. 2003, \apjl, 592, L53

\bibitem[{{van Dokkum} {et~al.}(2001){van Dokkum}, {Franx}, {Kelson}, \&
  {Illingworth}}]{2001ApJ...553L..39V}
{van Dokkum}, P.~G., {Franx}, M., {Kelson}, D.~D., \& {Illingworth}, G.~D.
  2001, \apjl, 553, L39

\bibitem[{{Verheijen}(2001)}]{2001ApJ...563..694V}
{Verheijen}, M.~A.~W. 2001, \apj, 563, 694

\bibitem[{{Willick}(1990)}]{1990ApJ...351L...5W}
{Willick}, J.~A. 1990, \apjl, 351, L5

\end{thebibliography}

\end{document}